\documentclass[twocolumn,showpacs,preprintnumbers,amsmath,amssymb]{revtex4}

\usepackage{graphicx}
\usepackage{dcolumn}
\usepackage{bm}

\begin{document}

\preprint{Submission to Phys. Rev. B}

\title{
Magnetic excitations of 
the spin-1/2 tetramer substance 
Cu$_2$$^{114}$Cd$^{11}$B$_2$O$_6$ 
obtained by inelastic neutron scattering experiments
}

\author{Masashi Hase$^1$}
 \email{HASE.Masashi@nims.go.jp}
\author{Kenji Nakajima$^2$}
\author{Seiko Ohira-Kawamura$^2$}
\author{Yukinobu Kawakita$^2$}
\author{Tatsuya Kikuchi$^2$}
\author{Masashige Matsumoto$^3$}


\affiliation{%
${}^{1}$National Institute for Materials Science (NIMS), 
Tsukuba, Ibaraki 305-0047, Japan \\
${}^{2}$J-PARC Center, 
Tokai, Ibaraki 319-1195, Japan \\
${}^{3}$Department of Physics, Shizuoka University, 
Shizuoka 422-8529, Japan
}%

\date{\today}

\begin{abstract}

We performed inelastic neutron scattering experiments 
on Cu$_2$$^{114}$Cd$^{11}$B$_2$O$_6$ 
powder.
The magnetic excitations at low temperatures
are similar to those of 
the interacting spin-1/2 tetramers in the ordered state. 
The weak excitations existing above 3 meV
suggest that 
the Higgs mode appears in Cu$_2$CdB$_2$O$_6$ 
at ambient pressure and zero magnetic field. 
We evaluated 
$J_1 = 27.3 \pm 1.0$ and $J_2 = -14.0 \pm 1.4$ meV 
for the intra-tetramer interactions and  
$J_3 = -0.4 \pm 0.2$ and $J_4 = 1.4 \pm 0.2$ meV 
for the inter-tetramer interactions. 
The spin gap in the isolated spin tetramer 
was calculated to be 1.6 meV, which  
is less than the effective inter-tetramer interaction value 
($3.6 \pm 0.8$ meV).
Therefore, antiferromagnetic long-range order is possible, although 
the ground state of the isolated spin tetramer is the spin-singlet state. 
We discuss the temperature dependence of the magnetic excitations. 

\end{abstract}


\maketitle

\section{INTRODUCTION}

Antiferromagnetic (AF) $XXZ$ models describe 
the competition between spin-singlet pairs and AF long-range order (AF-LRO).
The Hamiltonian is expressed as  
\begin{equation}
{\cal H} = 
\sum_{i,j} J_{ij} (\frac{S_{i+} S_{j-} + S_{i-} S_{j+}}{2} 
+ \Delta S_{iz} S_{jz}).
\end{equation}
The first term in the parenthesis stabilizes spin-singlet pairs, 
induces quantum fluctuation, and destroys AF-LRO,  
whereas the second term stabilizes AF-LRO. 

In spin-1/2 AF Heisenberg models ($\Delta = 1$) with 
nearest-neighbor exchange interactions, 
the dimensionality of the lattice affects the magnetism as follows. 
In a three-dimensional spin system on a simple cubic lattice, 
AF-LRO appears at finite temperatures. 
In a two-dimensional spin system on a square lattice, 
AF-LRO exists only at 0 K. 
The ground state (GS) is a combination of  
a N\'eel state and 
a resonating-valence-bond (RVB) state \cite{Anderson87}. 
The magnitude of ordered magnetic moments is reduced 
by the overlap of the spin-singlet RVB state.
In a one-dimensional spin system on a uniform chain, 
no AF-LRO exists even at 0 K.
The GS is the gapless spin-singlet state 
known as a Tomonaga-Luttinger liquid (TLL). 
The spin correlation in a TLL decays algebraically. 
A TLL is a quasi-long-range-ordered state and is at quantum criticality.
Therefore, infinitesimal interchain exchange interactions stabilize AF-LRO. 
Other perturbations such as the alternation of exchange interactions 
\cite{Hase93a,Hase93b,Hase93c} and 
next-nearest-neighbor exchange interactions 
\cite{Haldane82a,Haldane82b,Okamoto92}, 
on the other hand, 
generate a spin gap between the spin-singlet ground and first excited states 
and stabilize spin-singlet state(s).

In a zero-dimensional spin system on an isolated spin cluster, 
no AF-LRO exists even at 0 K. 
AF-LRO sometimes appears in weakly coupled spin clusters
even when the GS of the isolated spin cluster is the spin-singlet state. 
We understand the mechanism of the appearance of AF-LRO to be
as follows. 
The GS of weakly coupled spin clusters 
can be magnetic because of the following mechanism.
The values of $S^{\rm T}$ and $S^{\rm T}_z$ 
are 0 in the spin-singlet GS. 
Here, $S^{\rm T}$ and $S^{\rm T}_z$ represent  
the value and $z$ value, respectively, 
of the sum of the spin operators in a cluster. 
Other $S^{\rm T}_z  = 0$ states can be hybridized with 
the spin-singlet GS
of an isolated spin cluster by intercluster interactions \cite{Masuda09}.  
States with $S^{\rm T} > 0$ and $S^{\rm T}_z = 0$ are magnetic. 
For example, $S^{\rm T}_z$ is zero 
in a collinear two-sublattice AF ordered state, 
although the state is not an eigenstate of AF Heisenberg models. 
As a result, the GS of weakly coupled spin clusters 
can become magnetic by the hybridization of plural $S^{\rm T}_z = 0$ states. 

A gapless Nambu-Goldstone mode exists in AF-LRO
because of broken symmetries \cite{Goldstone62}. 
Therefore, the excitation energy of a mode must be zero 
below the transition temperature $T_{\rm N}$. 
Intercluster interactions change
the discrete energy level of an isolated cluster to 
an energy band with a finite width. 
As the temperature $T$ is lowered, 
the band width increases. 
If the spin gap 
is comparable to or less than the energy of 
the effective intercluster interaction, 
the gapless Nambu-Goldstone mode can appear and therefore 
AF-LRO can appear.   
Here, the effective intercluster interaction 
is given by the summation of 
the product of the absolute value of the intercluster interaction and 
the number of interactions per spin ($z$) \cite{Matsumoto10}.


There are several model substances with weakly coupled spin clusters
showing AF-LRO. 
The spin system in {\it A}CuCl$_3$ \ ({\it A} = NH$_4$, K, or Tl) 
consists of coupled Cu-Cu dimers. 
These substances can have AF-LRO even at zero magnetic field. 
NH$_4$CuCl$_3$ exhibits AF-LRO at ambient pressure 
\cite{Kurniawan99}, 
whereas KCuCl$_3$ and TlCuCl$_3$ 
exhibit AF-LRO above critical pressures 
\cite{Goto06,Tanaka03}. 
NH$_4$CuCl$_3$ has three types of dimers 
\cite{Matsumoto03,Ruegg04,Matsumoto15}. 
In one of them, 
the value of the intra-dimer interaction is 0.29 meV, which 
is less than the values of the effective inter-dimer interaction. 
Therefore, magnetic excitations generated by the dimer 
are gapless within the experimental accuracy \cite{Oosawa03}. 
The spin gap of KCuCl$_3$ is 2.60 meV, which 
is comparable to the value of the effective inter-dimer interaction 
\cite{Matsumoto04}. 
The spin gap of TlCuCl$_3$ is 0.70 meV, which
is less than the values of the effective inter-dimer interaction 
\cite{Matsumoto04}. 
As the pressure is raised or the temperature is lowered 
in TlCuCl$_3$, 
the triplet excitations are softened and 
become gapless at $T_{\rm N}$ 
\cite{Ruegg08,Merchant14}. 
In the ordered state, 
degenerate triplets are separated into 
transverse (Nambu-Goldstone) and longitudinal (Higgs) modes. 


Several AF-LROs can be explained 
in the scheme of weakly coupled spin clusters. 
For example, 
if we consider that 
the spin system in Cu$_2$Fe$_2$Ge$_4$O$_{13}$ 
comprises weakly coupled Fe-Cu-Cu-Fe tetramers (four-spin systems) 
\cite{Matsumoto10}, 
we can more easily understand 
the ordering mechanism of magnetic moments on Cu$^{2+}$ ions. 
The spin system was considered as 
a combination of spin-1/2 Cu dimers and spin-5/2 Fe chains 
\cite{Masuda09,Masuda03,Masuda04,Masuda05,Masuda07}. 
The values of the intradimer and intrachain exchange interactions 
($J_{\rm Cu}$ and $J_{\rm Fe}$, respectively) are 
22.0 and 1.60 meV, respectively.  
An exchange interaction ($J_{\rm Cu-Fe}$) 
couples a dimer and chain, where
the value of $J_{\rm Cu-Fe}$ is 2.30 meV. 
The magnetic excitations corresponding to 
singlet-triplet excitations in Cu dimers 
have been observed around 24 meV \cite{Masuda09}. 
In spite of the large value of $J_{\rm Cu}$, 
which appears to stabilize a (nearly) spin-singlet state of Cu spins, 
both Cu and Fe moments are cooperatively ordered 
below $T_{\rm N} = 39$ K. 
The magnitudes of Cu and Fe moments are 
0.38(4) and 3.62(3) $\mu_{\rm B}$, respectively, where 
the Cu moments are not so small. 

The Fe-Cu-Cu-Fe tetramer can be considered to be formed by 
the $J_{\rm Cu}$ and $J_{\rm Cu-Fe}$ interactions. 
The GS of the isolated tetramer is the spin-singlet state. 
The spin gap in the isolated tetramer was calculated to be 0.11 meV 
in spite of the large values of 
the $J_{\rm Cu}$ and $J_{\rm Cu-Fe}$ interactions. 
The spin gap is smaller than 
the value of the effective intercluster interaction.  
Accordingly, we can understand the appearance of the cooperative order 
in the scheme of weakly coupled spin clusters. 


The mechanism of the appearance of AF-LRO below $T_{\rm N} = 9.8$ K 
in Cu$_2$CdB$_2$O$_6$ \cite{Hase05,Munchau95}
appears to be analogous to that in Cu$_2$Fe$_2$Ge$_4$O$_{13}$. 
We observed a 1/2 quantum magnetization plateau 
above 23 T at 2.9 K.
A quantum magnetization plateau 
does not appear in conventional AF-LROs and  
indicates the existence of an energy gap. 
From the magnetization results \cite{Hase05} and 
the magnetic structure \cite{Hase09}, 
we have determined that 
the spin system in Cu$_2$CdB$_2$O$_6$ consists of  
weakly coupled spin-1/2 tetramers. 
Each tetramer is formed by 
the antiferromagnetic $J_1$ and 
ferromagnetic $J_2$ exchange interactions
as shown in Fig. 1(a). 
The values of  $J_1$ and $J_2$ were evaluated to be
264 K (22.7 meV) and -143 K (-12.3 meV) in the previous work \cite{Hase09}. 
The GS of the isolated tetramer is the spin-singlet state. 
The spin gap in the isolated tetramer was calculated to be 1.43 meV 
in spite of the large values of  the $J_1$ and $J_2$ interactions. 
In the previous papers \cite{Hase05,Hase09}, 
we considered only the $J_3$ interaction ($z=2$) as 
the inter-tetramer interaction. 
The $z|J_3|$ value was evaluated to be 9.9 K (0.85 meV), which 
is comparable to the spin gap. 
To investigate whether 
the AF-LRO in Cu$_2$CdB$_2$O$_6$ can be understood 
in the scheme of weakly coupled spin clusters,  
we performed inelastic neutron scattering (INS) experiments 
on Cu$_2$$^{114}$Cd$^{11}$B$_2$O$_6$ 
powder to obtain its magnetic excitations.  

\begin{figure}
\begin{center}
\includegraphics[width=7cm]{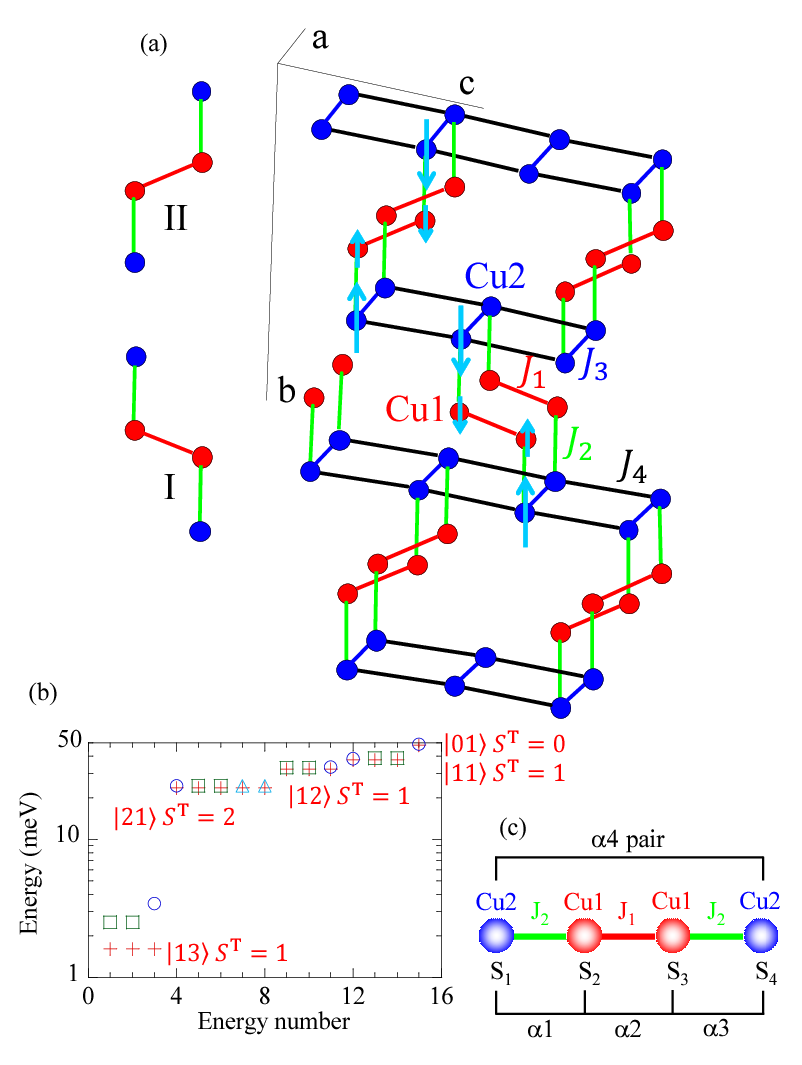}
\caption{
(Color online)
(a)
Schematic drawing of Cu$^{2+}$~ion positions having spin-1/2
in Cu$_2$CdB$_2$O$_6$. 
Two crystallographic Cu sites (Cu1 and Cu2) exist. 
Red and blue circles represent 
Cu1 and Cu2 sites, respectively. 
The $J_1$ and $J_2$ exchange interactions 
form spin tetramers. 
Two kinds of tetramers (I and II) exist, 
although the two are equivalent to each other 
as a spin system. 
We take the $J_3$ and $J_4$ inter-tetramer interactions
into account. 
Arrows indicate ordered magnetic moments below $T_{\rm N} = 9.8$ K. 
The space group is monoclinic $P2_1 /c$ (No. 14) \cite{Munchau95}.  
The lattice constants at 15 K are
$a=3.4047(5)$ \AA, $b=15.140(2)$ \AA, $c=9.298(1)$ \AA, and 
$\beta=92.80(1){}^{\circ}$ \cite{Hase09}. 
(b)
Energies of excited states from the GS in 
the isolated spin tetramer formed 
by the $J_1$ and $J_2$ interactions (+ symbol) 
and those in 
the tetramer in the ordered state (other symbols). 
We evaluated 
the exchange interaction parameters that  
reproduce the magnetic excitations observed 
by inelastic neutron scattering measurements. 
Their values are listed in Table I.
$S^{\rm T}$ is 
the value of the sum of the spin operators in the tetramer. 
The eigenstates $|ij \rangle$ of  the isolated tetramer
are explicitly given in Ref.~\cite{Hase97}. 
In the isolated tetramer, 
the GS is the spin-singlet $|02 \rangle$ state. 
Circles, squares, and triangles indicate 
states with $S^{\rm T}_z = 0, 1$, and 2, respectively, in the ordered state. 
(c) 
An illustration of the isolated tetramer 
to explain the $|02 \rangle$ and $|13 \rangle$ states. 
Details are described in the Appendix A. 
}
\end{center}
\end{figure}

\begin{table}
\caption{\label{table1}
Values of exchange interaction parameters (in the unit of meV)
obtained in the present study. 
We used the central values for calculations of 
the energies in Fig. 1(b), 
magnetic excitations in Figs. 4, 6(a), and 7(b). 
}
\begin{ruledtabular}
\begin{tabular}{cccc}
$J_1$ & $J_2$ & $J_3$ & $J_4$\\
\hline
$27.3 \pm 1.0$ & $-14.0 \pm 1.4$ & $-0.4 \pm 0.2$ & $1.4 \pm 0.2$\\
\end{tabular}
\end{ruledtabular}
\end{table}

\section{Experimental and Calculation Methods}

Crystalline Cu$_2$$^{114}$Cd$^{11}$B$_2$O$_6$ 
powder was synthesized by a solid-state reaction  
at 1,073 K in air for 160 h with intermediate grindings. 
We used the isotopes $^{114}$Cd and $^{11}$B 
to decrease the absorption of neutrons. 
The purity of the isotopes was 99 \%. 
We confirmed the formation of Cu$_2$CdB$_2$O$_6$ 
using an x-ray diffractometer (RINT-TTR III; Rigaku).
We performed INS measurements 
using the disk-chopper-type spectrometer (AMATERAS)
at BL 14 \cite{Nakajima11}
in the Materials and Life Science Experimental Facility (MLF)
of Japan Proton Accelerator Research Complex (J-PARC). 
We placed about 9 g of the powder 
in a vanadium cylinder with a diameter of 10 mm 
and mounted the cylinder 
in a $^4$He closed cycle refrigerator. 
The experimental data were obtained by 
the UTSUSEMI software provided by MLF 
\cite{Inamura13}. 

We considered the model shown in Fig. 1(a). 
The $J_1$ and $J_2$ interactions are intra-tetramer interactions. 
The $J_3$ and $J_4$ interactions are inter-tetramer  interactions. 
From the magnetic structure of Cu$_2$CdB$_2$O$_6$ \cite{Hase09}, 
the $J_1$ and $J_4$ interactions are antiferromagnetic and 
the $J_2$ and $J_3$ interactions are ferromagnetic. 
The ordered moments in the calculation are parallel to 
the corresponding ordered moments determined experimentally. 
The moments are nearly parallel to the $b$ axis. 
We calculated energies of ground and excited states of 
an isolated tetramer and a tetramer in the ordered state
using the exact diagonalization \cite{Hase97} and 
tetramer mean-field theory, respectively \cite{Matsumoto10}. 
We calculated the dispersion relation of magnetic excitations and 
the neutron scattering intensities of the model 
for Cu$_2$CdB$_2$O$_6$ shown in Fig. 1(a) 
using extended Holstein-Primakoff theory \cite{Matsumoto10}.

\section{Experimental results}


Figure 2 shows INS intensity $I(Q, \omega)$ maps of 
the Cu$_2$$^{114}$Cd$^{11}$B$_2$O$_6$ 
powder. 
Here, $Q$ and $\omega$ are 
the magnitude of the scattering vector and the energy transfer, respectively. 
The energy of incident neutrons $E_{\rm i}$ is 7.74 meV. 
Excitations at 5.3 K are most apparent around 
$\omega = 2.3$ meV and $Q = 0.5$ \ \AA$^{-1}$. 
The intensity of excitations between 1.7 and 2.9 meV 
is suppressed at higher $Q$. 
As the temperature $T$ is raised, 
the excitation energies are lowered and 
the intensity of the excitations is suppressed. 
These results mean that 
the observed excitations are dominated by those of magnetic origin. 
Weak excitations exist at $\omega \lesssim 1.5$ meV 
in the region $Q = 0.7 - 0.8$ \AA$^{-1}$ 
at 5.3 K.
The intensity of the excitations increases with $T$. 
Weak excitations also exist at $\omega \gtrsim 3.0$ meV at 5.3 K. 

\begin{figure}
\begin{center}
\includegraphics[width=8cm]{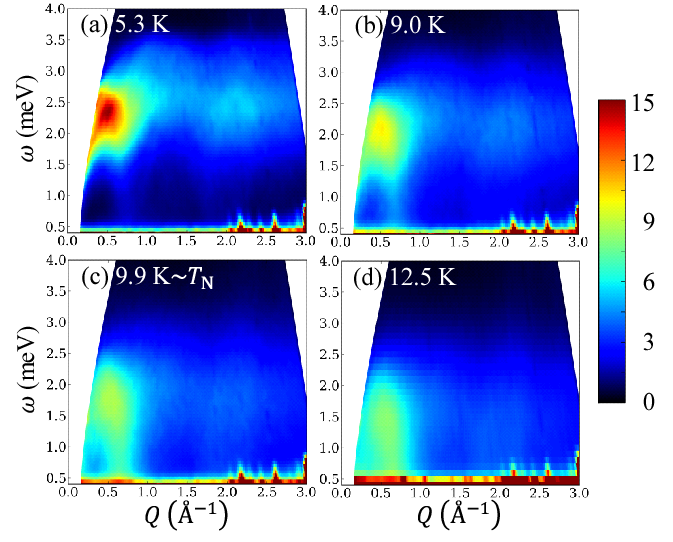}
\caption{
(Color online)
INS intensity $I(Q, \omega)$ maps 
in the $Q - \omega$ plane for the 
Cu$_2$$^{114}$Cd$^{11}$B$_2$O$_6$ 
powder 
at several temperatures. 
The energy of incident neutrons $E_{\rm i}$ is 7.74 meV. 
The right vertical key shows the INS intensity in arbitrary units. 
}
\end{center}
\end{figure}


Figure 3 shows the $\omega$ dependence of
${\rm Im} \chi (Q, \omega) \equiv I(Q, \omega) \ast 
(1-e^{-\omega / k_{\rm B} T})$ for the  
Cu$_2$$^{114}$Cd$^{11}$B$_2$O$_6$ 
powder. 
As shown in Fig. 4 later, 
the $Q$ dependence of the INS intensity has 
a maximum around $Q = 0.50$ and $0.55$ \ \AA$^{-1}$  
below and above $T_{\rm N} = 9.8$ K, respectively. 
Therefore, Figs. 3(a) and (b) show 
${\rm Im} \chi (Q, \omega)$ summed in the $Q$ ranges of 
$0.40 - 0.60$ and $0.45 - 0.65$ \AA$^{-1}$ 
below 9.0 K and above 9.9 K, respectively.
${\rm Im} \chi (Q, \omega)$ at 5.3 K shows 
a broad maximum whose center is at 2.3 meV. 
The horizontal bar indicates the energy resolution at $\omega = 0$ meV. 
${\rm Im} \chi (Q, \omega)$ at 5.3 K 
is broader than the energy resolution. 
As the temperature $T$ is raised, 
both the peak position as shown in the inset of Fig. 3 (b)
and the peak height are lowered. 
${\rm Im} \chi (Q, \omega)$ at $\omega$ below the peak energy 
increases with $T$ up to 12.5 K.
${\rm Im} \chi (Q, \omega)$ below 3.7 meV
decreases with increasing $T$ above 12.5 K. 
Figure 3(c) shows 
${\rm Im} \chi (Q, \omega)$ summed in the $Q$ range of 
$1.05 - 1.15$ \AA$^{-1}$ 
at 5.3 K. 
We can see a maximum around 2.5 meV. 
As explained later, 
we used the excitation energy to evaluate the $J_3$ value. 
 
\begin{figure}
\begin{center}
\includegraphics[width=8cm]{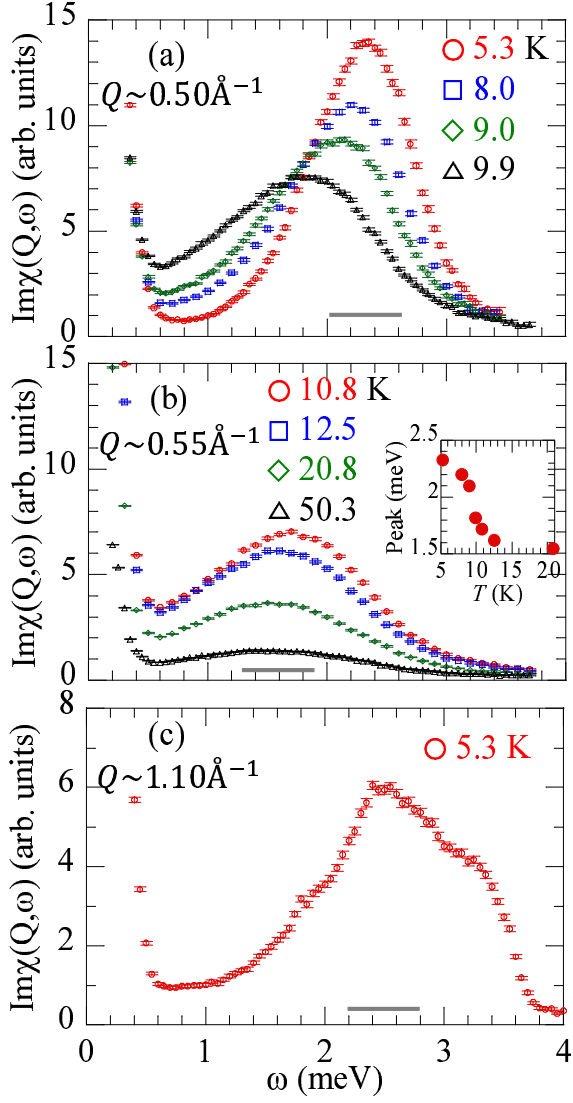}
\caption{
(Color online)
$\omega$ dependence of 
${\rm Im} \chi (Q, \omega) \equiv I(Q, \omega) \ast 
(1-e^{-\omega / k_{\rm B} T})$ for the
Cu$_2$$^{114}$Cd$^{11}$B$_2$O$_6$ 
powder. 
The energy of incident neutrons $E_{\rm i}$ is 7.74 meV. 
The horizontal bar indicates the energy resolution of 0.6 meV
at $\omega = 0$ meV. 
(a)
${\rm Im} \chi (Q, \omega)$ summed in the $Q$ range of 
$0.40 - 0.60$ \AA$^{-1}$ 
at $T \le 9.0$~K and 
$0.45 - 0.65$ \AA$^{-1}$ 
at 9.9 K. 
(b)
${\rm Im} \chi (Q, \omega)$ summed in the $Q$ range of 
$0.45 - 0.65$ \AA$^{-1}$ 
at $T \ge 10.8$~K. 
The inset shows the $T$ dependence of the peak energy 
of ${\rm Im} \chi (Q, \omega)$. 
(c)
${\rm Im} \chi (Q, \omega)$ summed in the $Q$ range of 
$1.05 - 1.15$ \AA$^{-1}$ 
at 5.3 K. 
}
\end{center}
\end{figure}


Figure 4 shows the $Q$ dependence of the INS intensity 
around the peak of the magnetic excitations in Fig. 3. 
The intensities were 
summed in the $\omega$ ranges of 
$2.20 - 2.40$ and $1.25 - 1.65$ meV 
at 5.3 and 12.5 K, respectively.
The intensity at 5.3 K in the ordered state 
shows broad peaks around 
$Q = 0.50$ and  $2.10$ \AA$^{-1}$. 
The intensity at 12.5 K in the paramagnetic state 
shows broad peaks around 
$Q = 0.55$ and  $2.10$ \AA$^{-1}$ 
and 
a shoulder around $Q = 1.20$ \ \AA$^{-1}$. 
We will explain the calculated lines later.

\begin{figure}
\begin{center}
\includegraphics[width=8cm]{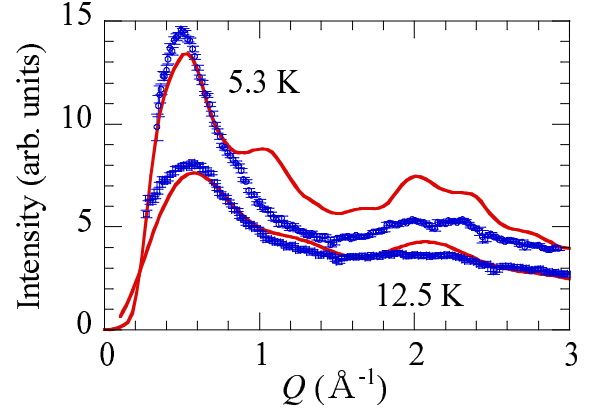}
\caption{
(Color online)
$Q$ dependence of the INS intensity for the 
Cu$_2$$^{114}$Cd$^{11}$B$_2$O$_6$ 
powder. 
The energy of incident neutrons $E_{\rm i}$ is 7.74 meV. 
The data show the INS intensity 
summed in the $\omega$ range of 
$2.20 - 2.40$ meV ($1.25 - 1.65$ meV) 
at 5.3 K (12.5 K).
The two lines indicate the calculated results described in the text.  
}
\end{center}
\end{figure}


Magnetic excitations with higher energies exist. 
Figure 5 shows the $\omega$ dependence of the INS intensity 
summed in the $Q$ range of
$1.8 - 2.2$ \AA$^{-1}$ 
for the Cu$_2$$^{114}$Cd$^{11}$B$_2$O$_6$ 
powder. 
The energy of incident neutrons $E_{\rm i}$ is 42.1 meV. 
Excitations can be seen at approximately 24 meV below $T_{\rm N}=9.8$ K. 
As $T$ is raised, 
the intensities around 24 meV decrease. 
Therefore, magnetic excitations exist around 24 meV. 
Excitations can be observed at approximately 22 meV above $T_{\rm N}$.  
As described later, we consider that 
magnetic excitations exist around 22 meV above $T_{\rm N}$. 
In addition to the magnetic excitations, 
other excitations can be seen in the energy range shown in Fig. 5. 
From the $T$ dependence, 
we consider that the other excitations are phonons.  

\begin{figure}
\begin{center}
\includegraphics[width=8cm]{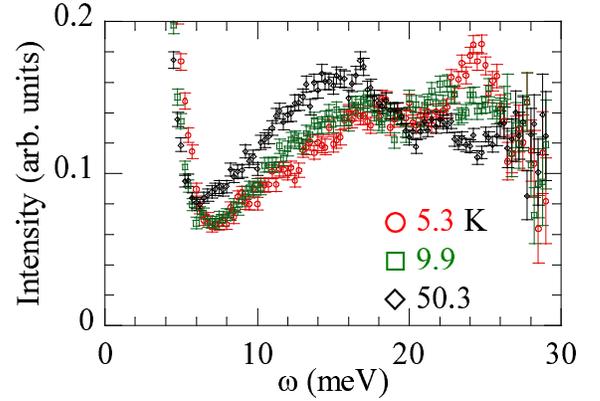}
\caption{
(Color online)
$\omega$ dependence of the INS intensity for the 
Cu$_2$$^{114}$Cd$^{11}$B$_2$O$_6$ 
powder. 
The energy of incident neutrons $E_{\rm i}$ is 42.1 meV. 
The data show the intensity 
summed in the $Q$ range of 
$1.8 - 2.2$ \AA$^{-1}$. 
Red circles, green squares, and black diamonds show 
the data at 5.3, 9.9, and 50.3 K, respectively.
}
\end{center}
\end{figure}

\section{Comparison between experimental and calculated results}


The values of the exchange interactions 
were evaluated as follows 
to explain the experimental results and 
are listed in Table I. 
We used the peak energies 
1.6 meV above $T_{\rm N}$ and 24 meV at 5.3 K 
to evaluate the $J_1$ and $J_2$ values. 
As described later, 
the inter-tetramer $J_3$ and $J_4$ interactions 
are less relevant above $T_{\rm N}$. 
The peak energy 1.6 meV is the energy difference of 
the GS and the first ESs of 
the isolated tetramer. 
The peak energy 24 meV corresponds to the energy difference of 
the GS and the ESs with $S^{\rm T} = 2$. 
In the ordered state at $T < T_{\rm N}$, 
the $|13 \rangle$ state 
participating in the GS enables the excitation.
The energy difference is nearly independent of 
the $J_3$ and $J_4$ values.
We evaluated $J_1 = 27.3 \pm 1.0$ and $J_2 = -14.0 \pm 1.4$ meV. 
We assumed that errors of  the peak energies 1.6 and 24 meV 
were 0.2 and 1.0 meV, respectively, and 
estimated the errors of  the $J_1$ and $J_2$ values. 
As described later, 
the strong intensities 
around $\omega = 2.3$ meV and  $Q = 0.5$ \ \AA$^{-1}$ 
are generated by the excitations of
the T$_0$ mode around $(0~1~0)$ ($Q = 0.42$ \ \AA$^{-1}$). 
The energy of the T$_0$ mode  
depends strongly on the $J_4$ value 
when the $J_1$ and $J_2$ values are fixed.
We evaluated $J_4 = 1.4 \pm 0.2$ meV. 
We assumed that the error of  the T$_0$ mode energy 
were 0.2 meV and 
estimated the error of  the $J_4$ value. 
As shown in Fig. 3(c), 
${\rm Im} \chi (Q, \omega)$ at 5.3 K around $Q = 1.1$ \ \AA$^{-1}$ 
shows the maximum at 2.5 meV.  
The excitation energy 
depends strongly on the $J_3$ value 
when the $J_1$ and $J_2$ values are fixed.
We evaluated $J_3 = -0.4 \pm 0.2$ meV. 
We assumed that the error of  the excitation energy 
were 0.2 meV and 
estimated the error of  the $J_3$ value. 
We confirmed that AF-LRO appeared 
in the spin model with evaluated values including the errors
of the exchange interactions.  


Figure 1(b) shows the energies of the excited states 
of the tetramer \cite{Matsumoto10,Hase97}. 
In the isolated tetramer, 
the GS is the spin-singlet $|02 \rangle$ state. 
The first excited states are the spin-triplet $|13 \rangle$ states
located at 1.6 meV.  
The next higher-energy excited states $|21 \rangle$
are located at 23.6 meV.  
In the tetramer in the ordered state, 
the GS and excited states around 3 meV 
consist mainly of 
$|02 \rangle$ and $|13 \rangle$ states of the isolated tetramer. 
The excited states are located at 
2.5, 3.4, 24.3 meV, and higher energies. 

Figure 6 shows 
the dispersion relations of the magnetic excitations and 
the neutron-scattering intensities 
calculated at 0 K for the model of Cu$_2$CdB$_2$O$_6$ shown in Fig. 1. 
The appearance of the magnetic moment below $T_{\rm N}$
lifts the degeneracy and 
the triplet excitation splits into a doublet and a singlet. 
The former and the latter  are accompanied 
by the transverse (T) and longitudinal (L) spin fluctuations 
of the ordered moment, respectively. 
They lead to 
the massless T modes (Nambu-Goldstone modes) and 
the massive L mode (Higgs mode), respectively,  
after taking account of the dispersion relations 
caused by the inter-tetramer interactions.
Each mode has two branches denoted by 
the subscripts "0" and "$Q$" (T$_0$, T$_Q$, L$_0$, and L$_Q$), 
reflecting the two (I and II) tetramers in a unit cell.

\begin{figure}
\begin{center}
\includegraphics[width=8cm]{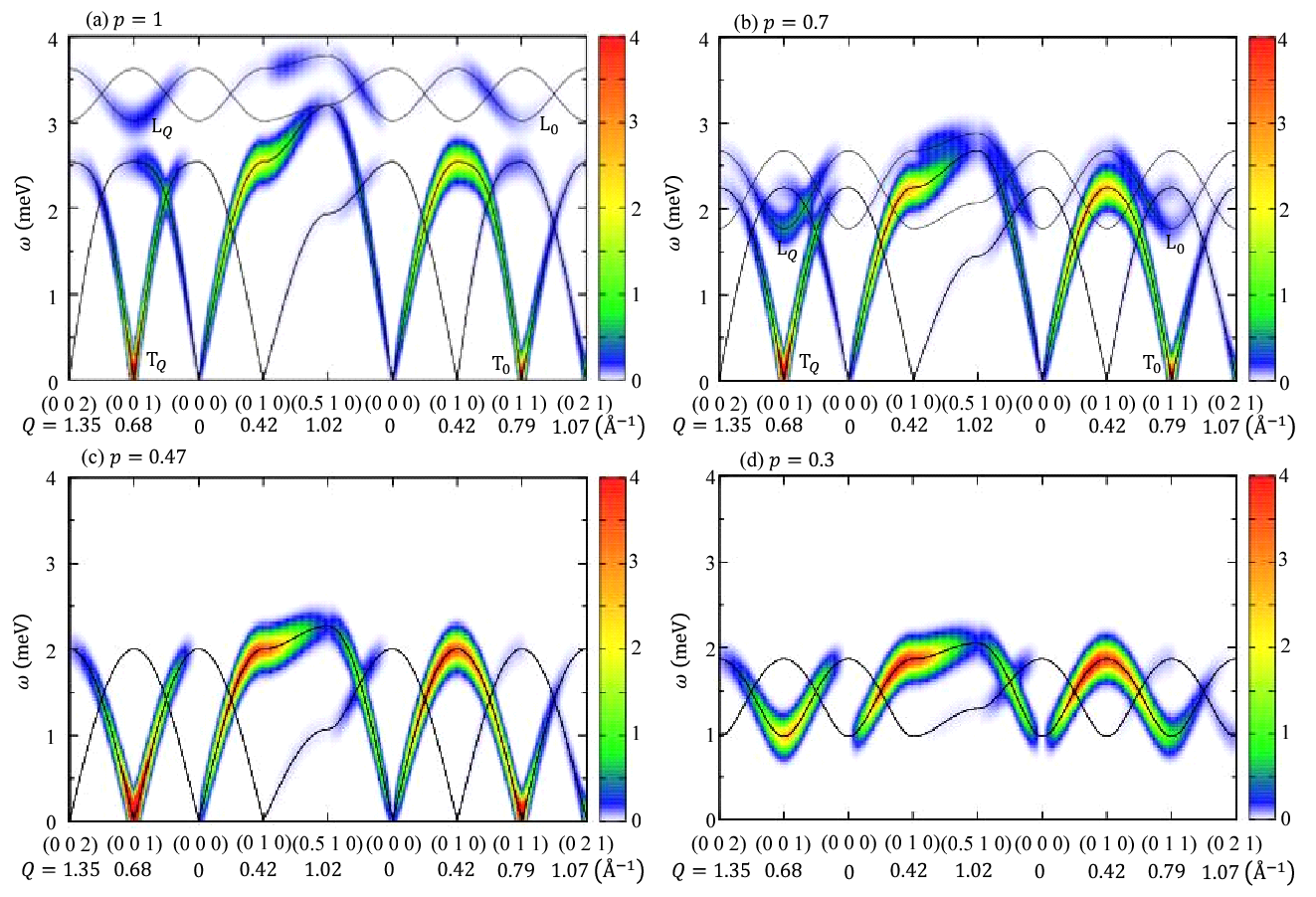}
\caption{
(Color online)
Dispersion relations of magnetic excitations (lines) 
along several symmetric axes 
calculated for the model of Cu$_2$CdB$_2$O$_6$ shown in Fig. 1 
using extended Holstein-Primakoff theory. 
The neutron-scattering intensities
are also depicted. 
The values of the exchange interactions are 
$J_1 = 27.3$, $J_2 = -14.0$, $J_3 = -0.4p$, and $J_4 = 1.4p$ meV 
with $p=1, 0.7, 0.47$, and $0.3$ 
in (a), (b), (c), and (d), respectively.  
We used a broadening factor of $\Gamma = 0.5$ meV 
for the Gaussian function
$\exp [-\frac{\{\omega - \omega_d ({\bf Q})\}^2}{\Gamma^2}]$, 
where 
$\omega_d ({\bf Q})$ is the excitation energy at ${\bf Q}$. 
In the ordered states (a) and (b), 
there are two transverse branches depicted as T$_0$ and T$_Q$, 
reflecting the two (I and II) tetramers in a unit cell, 
both of which are doubly degenerate. 
There are two L modes, L$_0$ and L$_Q$. 
In the critical state (c), 
the L and T modes become degenerate. 
The two lines reflect the two tetramers. 
In the paramagnetic state (d), 
the spin gap exists. 
}
\end{center}
\end{figure}


Figures 7(a) and 7(b) show 
the INS intensity map for 
the Cu$_2$$^{114}$Cd$^{11}$B$_2$O$_6$ 
powder at 5.3 K and 
that calculated for the model of Cu$_2$CdB$_2$O$_6$ shown in Fig. 1, 
respectively, in the $Q - \omega$ plane. 
The two maps are similar to each other. 
The intensities in both figures are strong
around $\omega = 2.3$ meV and  $Q = 0.5$ \ \AA$^{-1}$. 
The intensities are generated by the excitations of 
the T$_0$ mode around $(0~1~0)$ ($Q = 0.42$ \ \AA$^{-1}$). 
The weak excitations at $\omega \lesssim 1.5$ meV 
around $Q = 0.7 - 0.8$ \ \AA$^{-1}$ 
are caused by 
the branch  from $(0~1~1)$ ($Q = 0.79$ \ \AA$^{-1}$) 
of the T$_0$ mode and 
the branch  from $(0~0~1)$ ($Q = 0.68$ \ \AA$^{-1}$) 
of the T$_Q$ mode. 
These are the gapless Nambu-Goldstone modes 
that appear because of the magnetic order. 
The weak excitations at $\omega \gtrsim 3$ meV 
are caused by the excitations of 
the T$_0$ mode around $(0.5~1~0)$ ($Q = 1.02$ \ \AA$^{-1}$) 
and L mode. 
The Higgs mode (L mode) may appear in Cu$_2$CdB$_2$O$_6$ 
at ambient pressure and zero magnetic field. 
The spin gap was calculated to be 1.6 meV 
in the isolated tetramer with 
$J_1 = 27.3$ meV and $J_2 = -14.0$ meV.  
Consequently, we were able to confirm that 
the spin gap was less than 
the effective inter-tetramer interaction value 
($-2J_3+2J_4 = 3.6 \pm 0.8$ meV). 

\begin{figure}
\begin{center}
\includegraphics[width=8cm]{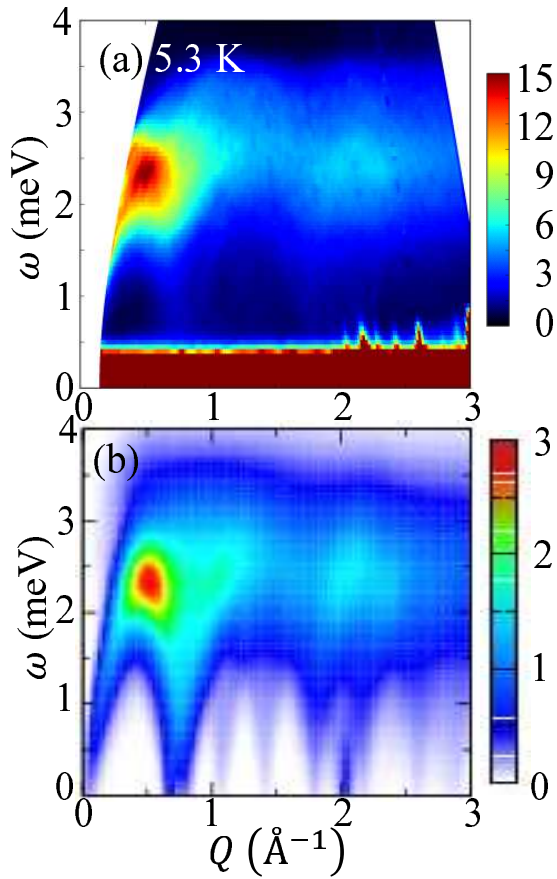}
\caption{
(Color online)
(a)
INS intensity $I(Q, \omega)$ map in the $Q - \omega$ plane for the 
Cu$_2$$^{114}$Cd$^{11}$B$_2$O$_6$ 
powder at 5.3 K. 
The energy of incident neutrons $E_{\rm i}$ is 7.74 meV. 
The right vertical key shows the intensity in arbitrary units. 
(b)
INS intensity map in the $Q - \omega$ plane 
calculated for the model of Cu$_2$CdB$_2$O$_6$ shown in Fig. 1 
using extended Holstein-Primakoff theory. 
The values of the exchange interactions are listed in Table I.
We used a broadening factor of $\Gamma = 0.5$ meV.
}
\end{center}
\end{figure}


We next compare the experimental and calculated  
$Q$ dependence of the intensity in Fig. 4. 
The upper line indicates the intensity of the excitation 
calculated for the model of Cu$_2$CdB$_2$O$_6$ shown in Fig. 1.  
The calculated intensities between 2.2 and 2.4 meV 
were summed. 
Although the intensities observed at 5.3 K are inconsistent with 
the upper line and 
the peak around $Q = 1.1$ \ \AA$^{-1}$ 
was not observed experimentally, 
the line is similar to the INS intensity at 5.3 K. 
The lower line indicates the intensity of the excitation 
from the spin-singlet GS 
to the spin-triplet first excited states 
in the isolated spin tetramer with $J_2 / J_1 = -0.51$. 
The line is close to the INS intensity at 12.5 K in the paramagnetic state. 
The peak position around $Q = 0.5$ \ \AA$^{-1}$ 
is slightly lower in the data at 5.3 K and upper line than 
in the data at 12.5 K and lower line, respectively. 
The width of the peak around $Q = 0.5$ \ \AA$^{-1}$ 
is slightly narrower in the data at 5.3 K and upper line than 
in the data at 12.5 K and lower line, respectively. 
The slight changes of the peak position and width suggest that 
the $J_3$ and $J_4$ interactions are relevant in the ordered state. 

\section{Discussion}

First, we consider the effects of temperature. 
In the spin dimer system TlCuCl$_3$, 
the thermally populated triplet excitations block their movement. 
This leads to the suppression of the effective inter-dimer interactions 
by increasing $T$ \cite{Ruegg08,Merchant14}. 
In Cu$_2$CdB$_2$O$_6$, 
the values of the inter-tetramer interactions 
($J_3 = -0.4 \pm 0.2$ and $J_4 = 1.4 \pm 0.2$ meV)
are comparable to or lower than the temperatures 
at which the data in Fig. 3 were obtained. 
The inter-tetramer interactions are less relevant at higher $T$ 
in the $T$ range in Fig. 3. 
Each tetramer is affected by internal magnetic fields 
generated by magnetic moments on neighboring tetramers.
As $T$ is raised, 
the magnitudes of the magnetic moments decrease and 
the effect of internal magnetic fields decreases. 
Accordingly, the excitation energies  
in weakly coupled tetramers 
decrease and approach those in the isolated tetramers. 

To investigate the effects of temperature, 
we calculated 
dispersion relations of the magnetic excitations at 0 K
for $J_1 = 27.3$, $J_2 = -14.0$, $J_3 = -0.4p$, and $J_4 = 1.4p$ meV
with $p \le 1$. 
Figure 6 shows the calculated results. 
The energy range where 
the magnetic excitations of the T modes exist 
decreases with $p$. 
The excitations around $Q = 0.5$ \ \AA$^{-1}$ 
in the experimental results 
are caused by the T$_0$ mode around $(0~1~0)$. 
The T$_0$ mode energy at $(0~1~0)$ 
decreases with $p$. 
The magnetic excitations of 
the L$_0$ mode around $(0~1~1)$ and the L$_Q$ mode around $(0~0~1)$ 
shift to markedly lower energies with decreasing $p$. 
The excitation gap of the L mode at $(0~1~1)$ and $(0~0~1)$ 
decreases accordingly and vanishes at the critical value $p = 0.47$ 
that corresponds to $T_{\rm N}$. 
The L and T modes become degenerate at $p = 0.47$. 
No ordered magnetic moment exists at $p = 0.3$. 
Excitations are gapped because of the paramagnetic state. 
The band width is narrower with decreasing  $p$.

The $p$ dependence is similar to 
the $T$ dependence of the experimental results. 
As $T$ is raised, 
the peak energy shifts to lower energy as shown in the inset of Fig. 3(b) 
and 
the intensity between 0.5 and 1.7 meV in Fig. 3(a) 
increases at $T < T_{\rm N}$.
Probably, 
we observed experimentally the suppression of the effective inter-dimer interactions 
by increasing $T$, 
although we could not prove experimentally the decrease of the band width 
with increasing $T$ at $T > T_{\rm N}$.

The inter-tetramer interactions also exist at $T > T_{\rm N}$ and 
generate excitation bands with finite widths 
as in the case that $p = 0.3$. 
The excitations from thermally excited states
in each band probably generated the continuous low-energy intensities in Fig. 3.
In addition, as described in the introduction, 
the GS is magnetic 
because of the inter-tetramer interactions.  
Paramagnetic scattering from the GS may
also contribute to the intensities at low $\omega$. 
Consequently, at $T > T_{\rm N}$, 
it is difficult to observe a clear excitation gap with powder samples, 
although the magnetic excitations are expected to be gapped 
because of the absence of gapless Nambu-Goldstone modes. 


The high-energy magnetic excitations in Fig. 5 
can also be explained by the spin tetramers. 
At $T > T_{\rm N}$, 
the excitations around 22 meV correspond to 
those between the 
$|13 \rangle (S^T=1)$ and $|21 \rangle (S^T=2)$ states, which have  
an energy difference of 22.0 meV in the isolated spin tetramer. 
Note that the excitation from the GS $|02 \rangle (S^T=0)$
to the $|21 \rangle (S^T=2)$  (23.6 meV) state 
is forbidden in the isolated spin tetramer. 
In the ordered state at $T < T_{\rm N}$, 
the excitations around 24 meV correspond to those 
from the ground state to the excited states 
located at 24.3 meV as shown in Fig. 1(b). 
In this case, the $|13 \rangle$ state 
participating in the GS enables the excitations.

The origin of the 1/2 quantum magnetization plateau 
is essentially the energy difference between the 
$|13 \rangle (S^T=1)$ and $|21 \rangle (S^T=2)$ states. 
The latter state cannot contribute to 
the magnetization because of the energy difference. 
The magnetization cannot increase from $S^T_z =1$ and 
the 1/2 quantum magnetization plateau  
continues until 
one of $|21 \rangle$  states with $S^T_z =2$ 
is stabilized by the magnetic field.
We calculated the magnetization curve at 0 K 
using the extended Holstein-Primakoff (EHP) theory and 
that at 2.9 K 
using a quantum Monte Carlo (QMC) technique (not shown). 
The 1/2 magnetization plateau appears 
above $H_{\rm p} = 33$, 42, and 23 T 
in the EHP, QMC, and experimental results, respectively. 
We could not find $J$ values 
which could explain both the INS and magnetization results. 
The $H_{\rm p}$ value depends strongly on the $J_4$ interaction. 
We expect that 
both the INS and magnetization results 
can be reproduced by calculated results including 
other weak inter-tetramer interactions.  
To solve the discrepancy between the $H_{\rm p}$ values, 
we need more precise information on exchange interactions 
that may be obtained in INS experiments using single crystals 
of Cu$_2$CdB$_2$O$_6$. 
We will be able to confirm experimentally 
the existence of the Higgs mode in INS experiments using single crystals.  


The magnetic excitations around 24 meV in Fig. 5 
indicate that 
the character of the spin cluster 
(energy gap) 
remains even in the ordered state. 
This is probably a common feature of spin clusters. 
As described earlier, 
in Cu$_2$Fe$_2$Ge$_4$O$_{13}$, 
magnetic excitations around 24 meV generated by Cu dimers 
were observed below $T_{\rm N}$ \cite{Masuda09}. 
The spin system of Cu$_3$Mo$_2$O$_9$ 
consists of AF chains and dimers \cite{Hamasaki08,Hase08}. 
Magnetic excitations around 6 meV mainly generated by the dimers 
were observed below $T_{\rm N}$ \cite{Kuroe11,Matsumoto12}. 
AF-LRO appears upon the substitution of other ions for Cu or Ge sites 
in the spin-Peierls substance CuGeO$_3$ 
\cite{Hase93a,Hase93b,Hase93c}.
Magnetic excitations originating from the singlet-triplet excitation of dimers 
remain below $T_{\rm N}$ \cite{Martin97}. 
In the spinel antiferromagnets ZnCr$_2$O$_4$ and MgCr$_2$O$_4$, 
quasielastic scattering in the paramagnetic state can be explained 
by the spin-molecule (hexamer) model \cite{Lee02,Tomiyasu08}, where
the spin molecules are generated 
by geometrical frustration. 
Two modes appear at 4.5 and 9.0 meV in the ordered state and 
can be explained by hexamers and heptamers, 
respectively \cite{Tomiyasu08}. 
The spin molecules are interpreted to be quasiparticles of 
highly frustrated spins.  



Finally, we comment on another spin model for Cu$_2$CdB$_2$O$_6$ 
proposed in Ref~\cite{Janson12}.
By performing extensive density functional theory band-structure calculations,
four dominant exchange interactions were identified. 
The four exchange interactions 
form a frustrated quasi-two-dimensional magnetic model. 
This spin model accounts for the magnetization results. 
It is important to investigate theoretically whether 
the spin model can also explain the magnetic excitations 
observed in our study. 

\section{Conclusion}

We performed inelastic neutron scattering experiments 
on Cu$_2$$^{114}$Cd$^{11}$B$_2$O$_6$ 
powder.
The magnetic excitations at low temperatures
are similar to those of 
the spin-1/2 tetramers in the ordered state. 
The weak excitations at $\omega \gtrsim 3$ meV 
suggest that 
the Higgs mode (L mode) appears in Cu$_2$CdB$_2$O$_6$ 
at ambient pressure and zero magnetic field. 
We evaluated
$J_1 = 27.3 \pm 1.0$ and $J_2 = -14.0 \pm 1.4$ meV 
for the intra-tetramer interactions and  
$J_3 = -0.4 \pm 0.2$ and $J_4 = 1.4 \pm 0.2$ meV 
for the inter-tetramer interactions. 
With these exchange interaction parameters, 
the low-energy excitations of Cu$_2$CdB$_2$O$_6$ 
are dominated by the singlet-triplet states of a tetramer. 
This means that 
the low-energy physics can be described by 
an interacting spin-dimer (singlet-triplet) system.
The spin gap in the isolated spin tetramer 
was calculated to be 1.6 meV, 
which is less than the effective inter-tetramer interaction value 
($3.6 \pm 0.8$ meV).
Therefore, antiferromagnetic long-range order is possible, although 
the ground state of the isolated spin tetramer is the spin-singlet state. 
As the temperature $T$ is raised, 
the magnetic excitations shift to lower energies and 
the intensities at low energies increase. 
The temperature dependences can be understood as resulting from  
the less relevant inter-tetramer interactions at higher $T$,  
the decrease of the magnetic moments, and 
the decrease of the excitation energies of 
the longitudinal mode (Higgs mode)  
around the gapless points. 
Consequently, as $T$ is raised, 
the excitation energies in weakly coupled tetramers 
approach those in the isolated tetramers. 
The spin gap in the isolated spin tetramer (1.6 meV)
is close to the peak position of the magnetic excitations 
above $T_{\rm N}$. 

\begin{acknowledgments}

This work was partially supported 
by KAKENHI (No. 23540396) and grants 
from National Institute for Materials Science (NIMS). 
M. M. was supported by KAKENHI (No. 26400332).
The neutron scattering experiments were approved by 
the Neutron Science Proposal Review Committee of 
Japan Proton Accelerator Research Complex (J-PARC)/
Materials and Life Science Experimental Facility (MLF)
(Proposal No. 2013A0008) and 
supported by the Inter-University Research Program on 
Neutron Scattering of 
Institute of Materials Structure Science (IMSS), 
High Energy Accelerator Research Organization (KEK).
We are grateful  
to M. Kohno, T. Masuda, H. Kuroe, and K. Tomiyasu
for fruitful discussions and 
to S. Matsumoto 
for sample syntheses and x-ray diffraction measurements.   

\appendix
\section{$|02 \rangle$ and $|13 \rangle$ states}

We explain the $|02 \rangle$ and $|13 \rangle$ states
of the isolated tetramer
using Fig. 1(c). 
We designate a spin pair formed by 
$S_j$ and $S_{j+1}$ as an $\alpha_j$ pair. 
The $|02 \rangle$ state is expressed as 
$
C_{01}
(|\downarrow \downarrow \uparrow \uparrow \rangle 
+ | \uparrow \uparrow \downarrow \downarrow \rangle 
- |\downarrow \uparrow \downarrow \uparrow \rangle 
- |\uparrow \downarrow \uparrow \downarrow \rangle) + 
C_{02}
(|\downarrow \uparrow \downarrow \uparrow  \rangle 
+ | \uparrow \downarrow \uparrow \downarrow \rangle 
- |\uparrow \downarrow \downarrow \uparrow  \rangle 
- |\downarrow \uparrow \uparrow \downarrow \rangle) =
C_{01}
(|\downarrow \uparrow \rangle - 
|\uparrow \downarrow \rangle)_{\alpha 2} * 
(|\downarrow \uparrow \rangle 
- |\uparrow \downarrow \rangle)_{\alpha 4} +
C_{02}
(|\downarrow \uparrow  \rangle 
- |\uparrow \downarrow \rangle)_{\alpha 1} * 
(|\downarrow \uparrow \rangle 
- |\uparrow \downarrow \rangle)_{\alpha 3}.
$
The symbol $\uparrow$ and $\downarrow$ in kets $|.... \rangle$ 
means 
$S_{jz} = 1/2$ and -1/2 ($j = 1$ to 4), respectively. 
For example, $|\downarrow \downarrow \uparrow \uparrow \rangle$ means 
$|S_{1z}=-1/2, S_{2z}=-1/2, S_{3z}=1/2,S_{4z}=1/2 \rangle$.
The first term is a product of 
a singlet state in the $\alpha_2$ pair and a singlet state in the $\alpha_4$ pair.   
The second term is a product of 
a singlet state in the $\alpha_1$ pair and a singlet state in the $\alpha_3$ pair.   
The two coefficients are 
$C_{01} = \frac{1}{\sqrt{ 3 + (-1 + 4j + 2 \sqrt{1 - 2j + 4j^2})^2}}$ and 
$C_{02} = \frac{2 -4j -2 \sqrt{1 - 2j + 4j^2}}{2 \sqrt{ 3 + (-1 + 4j + 2 \sqrt{1 - 2j + 4j^2})^2}}.$
The $|13 \rangle$ state with $S^{\rm T}_z=1$ is expressed as 
$
C_{11}
(|\uparrow \downarrow \uparrow \uparrow \rangle 
- |\uparrow \uparrow \downarrow \uparrow \rangle) + 
C_{12}
(|\downarrow \uparrow \uparrow \uparrow \rangle 
- |\uparrow \uparrow \uparrow \downarrow  \rangle)= 
C_{11}
(|\downarrow \uparrow \rangle 
- |\uparrow \downarrow \rangle)_{\alpha 2}* 
|\uparrow \uparrow  \rangle_{\alpha 4} +
C_{12}(|\downarrow \uparrow \rangle 
- |\uparrow \downarrow \rangle)_{\alpha 4}* 
|\uparrow \uparrow \rangle_{\alpha 2}.
$
The first term is a product of 
a singlet state in the $\alpha_2$ pair and 
a triplet state with $S_z = 1$ in the $\alpha_4$ pair.   
The second term is a product of 
a singlet state in the $\alpha_4$ pair and 
a triplet state with $S_z = 1$ in the $\alpha_2$ pair.   
The two coefficients are  
$C_{11} = \frac{1 + j + \sqrt{1 + j^2}}{2 \sqrt{1 + (j + \sqrt{1 + j^2})^2}}$ 
and 
$C_{12} = \frac{1 - j - \sqrt{1 + j^2}}{2 \sqrt{1 + (j + \sqrt{1 + j^2})^2}}.$

\end{acknowledgments}

\newpage 

\end{document}